\documentstyle[prl,aps,preprint]{revtex} 
\begin{document}
\tightenlines
\preprint{\hbox to \hsize{
\hfill$\vcenter{\hbox{\bf ULB-TH/01-08}
\hbox{\bf MADPH-01-1221}
\hbox{April 2001}}$ }}

\title{\vspace{.35in}%
The prompt TeV-PeV atmospheric neutrino window}
\vspace{.1in}
\author{C.G.S. Costa$^{(1)}$,  
F. Halzen$^{(2)}$ and C. Salles$^{(1)}$}
\address{$^{(1)}$Service de Physique Th\'eorique, CP 225, 
Universit\'e  Libre de Bruxelles, 
Blvd du Triomphe, 1050 Brussels, Belgium}
\address{$^{(2)}$Department of Physics,
University of Wisconsin, Madison, WI 53706, USA}

\maketitle
\begin{abstract}
We discuss the possible existence of an observational window, in
the TeV-PeV energy range, for the detection of prompt neutrinos
from the decay of charmed particles produced in cosmic ray
interactions with the atmosphere. We calculate the event rates for
muon and tau neutrinos of heavy quark, mostly charm, origin. We argue  
that their
prompt fluxes are observable in a kilometer-scale neutrino telescope,
even though the calculations are subjected to large uncertainties,
which we quantify. 
We raise the possibility that a small component of prompt
neutrinos may already be present in the observed samples of current
experiments. We also discuss the interplay of the predicted fluxes
with those produced by the flavor oscillation of conventional
atmospheric neutrinos, and by anticipated cosmic sources.
\end{abstract}


%
The quest to search for sources of cosmic neutrinos beyond the sun, to  
search for the particles that constitute the cold dark matter, and to  
exploit other science opportunities ranging from astronomy to particle  
physics, led to the commissioning of large volume high-energy neutrino  
telescopes. While AMANDA and Baikal are taking 
data\cite{amanda,baikal}, similar and much larger instruments are
contemplated\cite{antares,nestor,ice3,nemo}. 
These detectors collect  
the Cherenkov radiation emitted by charged secondaries
(electrons, muons and taus) produced in neutrino charged
current interactions in the ice or water surrounding the optical
sensors. In order to filter the background of atmospheric muons
created by cosmic-ray   interactions, only upward-going neutrinos
which traverse the Earth are monitored. The first mission of a
new neutrino telescope is to calibrate the detector on the known
flux of atmospheric neutrinos\cite{nature}. Up to about 10\,TeV,
the main source of atmospheric neutrinos is the decay of pions
and kaons in the atmosphere produced in the interactions of
cosmic rays with the Earth's atmosphere; we will refer to them
as constituting the ``conventional'' atmospheric neutrino flux.
At higher energies, these mesons interact rather than decaying
into a neutrino because of the increasing lifetime of the parent
mesons. Therefore the semileptonic decay of very-short lived
charmed particles becomes the dominant atmospheric source, 
giving rise to the ``prompt'' neutrino flux.
The energy dependence of prompt neutrinos follows the cosmic ray
spectrum whereas the spectrum of conventional neutrinos is steeper by
one power in energy because of the competition of decay with
interaction of the parent particles. The prompt neutrino flux is
independent of zenith angle whereas conventional neutrinos are
preferentially produced in the rarified atmosphere at large zenith.

In this letter we discuss the possible existence of an observational  
window for prompt neutrinos in the TeV-PeV energy range. We calculate  
neutrino induced muon and tau event rates, recognizing that the fluxes  
are subject to large uncertainties arising from the combination of  
extreme atmospheric cascade parameters with different charm production  
models. We will argue that the prompt fluxes are observable in  
high-energy neutrino telescopes and analyse the interplay of the  
predicted fluxes with those produced by the flavor oscillation of  
conventional atmospheric neutrinos, and by anticipated cosmic sources.

Prompt neutrinos are produced by production and semileptonic decay of  
heavy quarks produced in cascades initiated by cosmic rays in the  
atmosphere. In order to appreciate the main ingredients in calculating  
the production of prompt leptons in the atmosphere, we write
an approximate solution to the cascade equations\cite{prs:99},
valid for energies below the charm critical energy ($\approx
10^{8}$~GeV):
\begin{equation}
\Phi_{l} (E) = Z_{pi}(E) \; Z_{il}(E) \;
\frac{\Lambda_{p}(E)}{ \lambda_{p}(E)}
\: \Phi_{p} (E,0),
\label{eq:numu}
\end{equation}

\noindent where $\Phi_{l} (E)$ is the flux of secondary leptons  
($l=\mu,\nu_{e}$ and $\nu_{\mu}$) from the decay of charmed hadrons  
($i= D, D_{s}$ or $\Lambda_{c}$) produced by cosmic ray hadrons. The  
charm production spectrum-weighted moment $Z_{pi}(E)$ describes the  
inclusive charm production cross section; $Z_{il}(E)$ is the charm  
decay spectrum-weighted moment which represents the semileptonic  
three-body decay kinematics; $\Lambda_{p}(E)$ and $\lambda_{p}(E)$ are,  
respectively, the proton attenuation and interaction lengths in  
air\cite{gaisser:92}. The primary cosmic ray flux at the top of the  
atmosphere $\Phi_{p} (E,0)$ is assumed to be composed of protons  
although the validity of the superposition approximation guarantees  
that the calculation extends to heavier nuclei. Above the charm  
critical energy (the energy where interaction and decay of the charm  
particle compete in the evolution of the cascade), the charm  
interaction length $\lambda_{i} (E)$ must also be taken into account  
and the complete solution has to be considered. What the prompt  
tau-neutrino flux ($l=\nu_{\tau}$) is concerned, it suffices to  
calculate a generalized decay Z-moment $Z_{il}(E)$, in  
Eq.~(\ref{eq:numu}) which accounts for the decay chain: $D_{s}  
\rightarrow \tau \nu_{\tau}$ followed by  $\tau \rightarrow \nu_{\tau}  
X$ (where $X$ is either a $\mu \nu_{\mu}$ pair or a meson)\cite{pr:99}.  

Detailed investigation of the charm induced muon-neutrino  
flux\cite{cookbook} reveals the large uncertainties in the prompt  
component resulting from the imprecise knowledge of atmospheric  
particle showering parameters: primary cosmic ray spectrum $\Phi_{p}$,  
spectrum-weighted moments $Z_{i\nu}$, attenuation and interaction  
lengths, $\Lambda_{p}$, $\lambda_{p}$ and $\lambda_{i}$. Further  
uncertainty is associated with the extrapolation to high energy of a  
variety of models describing the accelerator data on charm production  
such as the Quark Gluon String Model (QGSM)\cite{qgsm}, Recombination  
Quark Parton Model (RQPM)\cite{rqpm} and perturbative QCD  
(pQCD)\cite{pqcd}. In the end, we conclude that the calculated fluxes,  
for both $\nu_{\mu}$ and $\nu_{\tau}$, are subjected to uncertainties  
of up to one order of magnitude for a given charm model. When  
comparison is made of alternative charm production models, the spread  
reaches two orders of magnitude. We will therefore define an allowed  
range between the maximum and minimum prompt neutrino fluxes.

There are two contrasting approaches in calculating charm production  
cross sections. The first is to blindly apply perturbative QCD. This is  
a questionable procedure because, at the extreme energies considered  
here, large logarithms connected with the small ratio of the charm  
quark mass and the square root of the center of mass energy, $x = m_c /  
\sqrt{s}$, are likely to modify the calculation. For instance,
this procedure, even though subject to large ambiguities,
routinely underestimates the production of strange particles at
accelerator energies. It may therefore be more appropriate to
calculate charm production in the context of QCD-inspired models
that have been patterned to accommodate accelerator data on
strange particle production. QGSM and RQPM fall into this
category and predict similar and larger extrapolations for the
charm production cross section at higher energies.

The detection rates of upward-going prompt neutrinos, i.e. the rates  
of neutrino induced muons and taus, per year per effective detection  
area (in $2\pi$ sr), can be estimated from our flux calculations  
following the prescription of Gandhi {\it et al.}\cite{Gandhi:98}. We  
first concentrate on the operating neutrino telescopes, BAIKAL  
NT-200\cite{baikal} and AMANDA-II\cite{amanda}, and subsequently  
consider the proposed km$^{3}$ experiment, ICECUBE\cite{ice3}; see  
Table~\ref{table:1}. Calculations use the maximum and minimum allowed  
prompt fluxes. We show for comparison the expected rates for  
muon-neutrinos from conventional atmospheric decays. Absorption in the  
Earth of the highest energy neutrinos is taken into account. From  
Table~\ref{table:1} we anticipate a situation where more then a  
thousand prompt muon-neutrinos above 1~TeV may be extracted from  
ICECUBE data using their flat zenith angle distribution as a signature.  
Furthermore, we predict about 200 prompt tau-neutrino candidates which  
also produce characteristic showers in the  
detector\cite{stanev:prl83}. Above 10~TeV, the 20 remaining  
tau-neutrinos will have no counterpart from the tau-neutrinos resulting  
from the oscillation of conventional atmospheric neutrinos, a source  
of which we consider further on.

Our calculations raise the possibility that prompt neutrinos may be  
observed by high-energy neutrino telescopes, especially because models  
extrapolating to the larger fluxes are, at least in our opinion, more  
robust. We next discuss the interplay of the predicted fluxes with  
those produced by the flavor oscillation of conventional atmospheric  
neutrinos, and by anticipated sources of cosmic neutrinos.

\begin{figure}
\begin{center}
\vspace*{14cm}
\hbox{\hspace{-1.5cm}
        \includegraphics{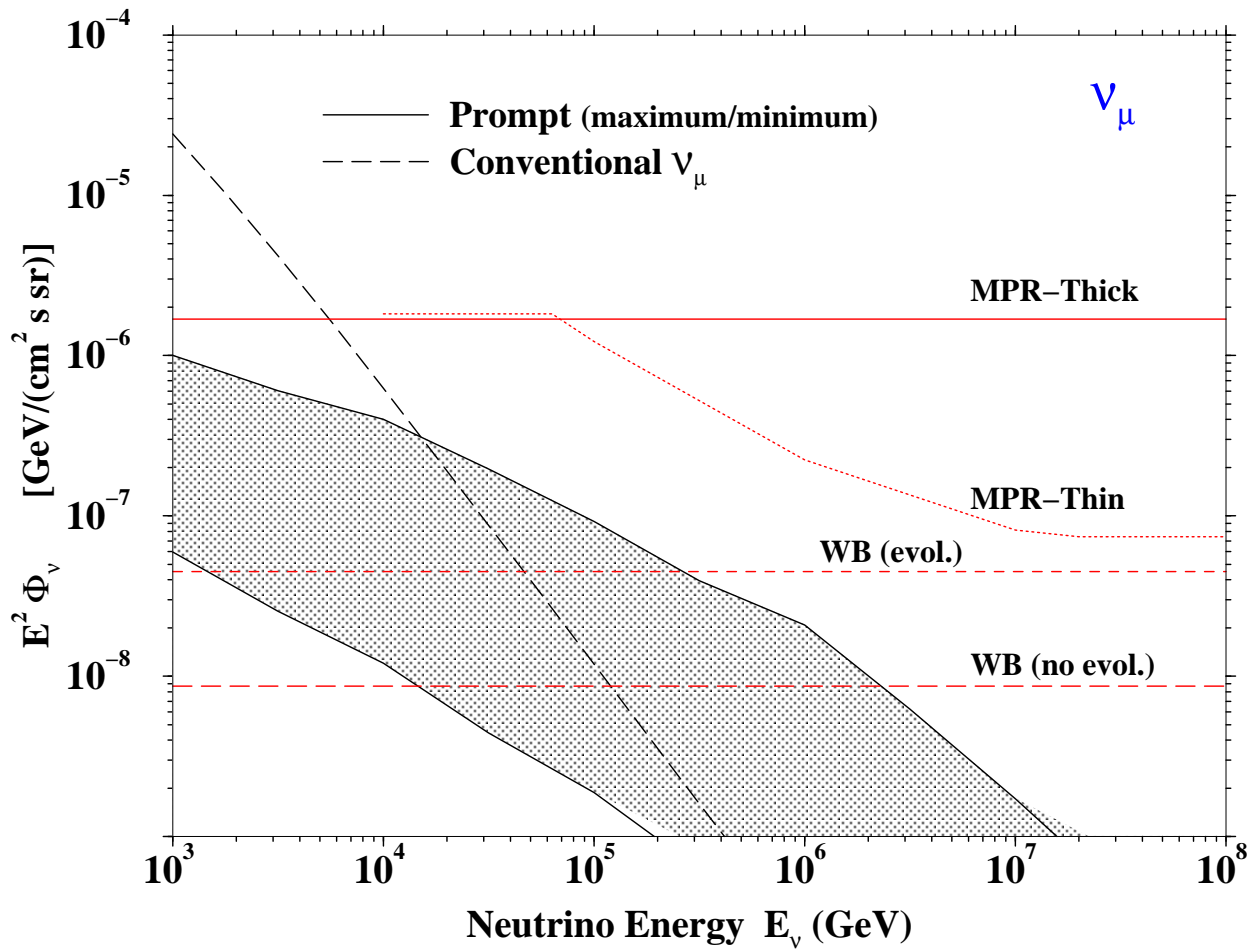}}
\end{center}
\caption[]{Comparison of several contributions to
the high-energy $\nu_{\mu}$ flux. The band represents the
allowed range between  maximum and minimum atmospheric 
prompt neutrino fluxes. Thick dashed line is the conventional 
atmospheric flux.
Solid, dotted, dashed and long-dashed lines correspond to upper
bounds imposed to diffuse extragalactic neutrino flux by the
observation of high-energy cosmic-ray and gamma-ray spectra,
according to different scenarios explained in the text.
Fluxes are multiplied by $E^{2}$.}
\label{fig:1}
\end{figure}

In Figure~\ref{fig:1} we present prompt $\nu_{\mu}$ fluxes  
corresponding to the allowed range between maximum and minimum 
predictions\cite{cookbook}, along with the conventional
atmospheric flux\cite{pqcd}. Among the potential candidates of
other high-energy neutrino sources\cite{reports}, we will
concentrate on cosmic accelerators such as active galactic nuclei
and gamma ray bursts which produce the highest energy photons
observed, and are theorized to produce the highest energy cosmic
rays. Their   anticipated diffuse neutrino flux is bounded by
experimental   information on measured high-energy fluxes of
cosmic rays, specifically neutrons, and gamma rays. The two upper
curves in Figure~\ref{fig:1} labeled MPR
(solid and dotted lines),
represent the allowed range of neutrino fluxes from sources that
are optically thick or thin to the emission of
neutrons\cite{mpr}, respectively. The two intermediate straight
lines labeled WB (dashed and long--dashed), represent bounds  
imposed on sources which produce the highest energy cosmic  
rays\cite{wb}. The higher flux allows for cosmological evolution
of the sources.

We note that the energy where the prompt neutrino flux exceeds the  
conventional one is roughly 10~TeV. Separation of the two components  
can be made, taking advantage of the fact that the conventional  
atmospheric flux has a characteristic and calculable zenith angle  
distribution, while the prompt component is isotropic.

If cosmic sources exist near the MPR bounds, the prompt muon-neutrino  
window is definitely closed. Neutrino telescopes will detect cosmic  
neutrinos at very high rates and the prompt component is a background  
to be dealt with. The AMANDA experiment is already exploring this range  
of fluxes\cite{nature}. Between $10^{5}$~GeV and  $10^{6}$~GeV the MPR  
bound for thin sources has essentially the same slope as the prompt  
flux. Since the bound is an upper limit, it may happen that the cosmic  
sources actually produce as many detectable neutrinos as atmospheric  
interactions do, and it will be difficult to disentangle both  
components. The situation is quite different when considering the  
extragalactic flux subjected to the WB-bounds. For evolving sources,  
there is room for an excess of prompt neutrinos in the range  
20-300~TeV. For non-evolving sources scenario, then the prompt window  
may be open up to about 2~PeV.

\begin{figure}
\begin{center}
\vspace*{14cm}
\hbox{\hspace{-1.5cm}
        \includegraphics{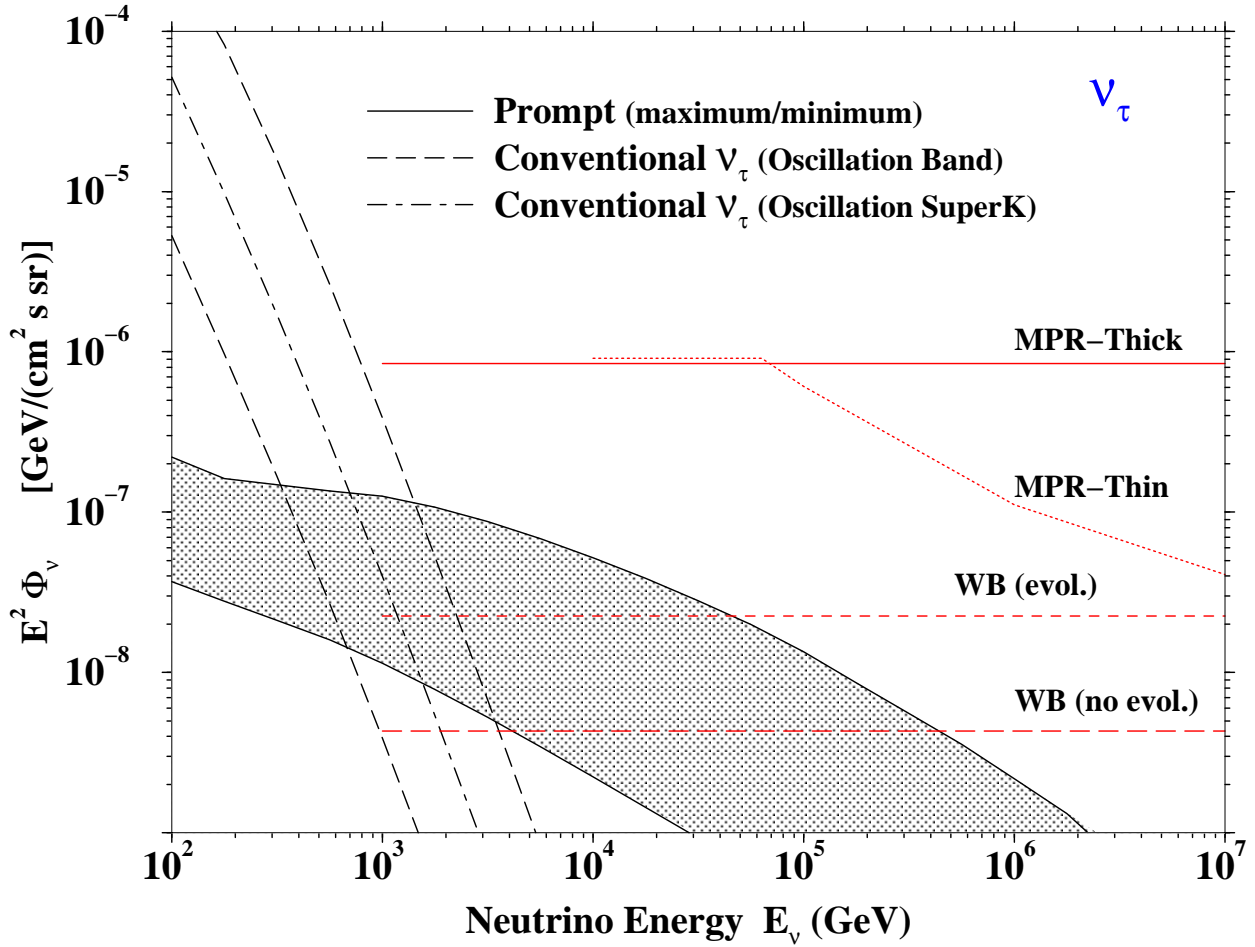}}
\end{center}
\caption[]{Comparison of several possible contributions
to the high-energy $\nu_{\tau}$ flux. The band represents the
allowed range between  maximum and minimum atmospheric 
prompt neutrino fluxes.
Thick dashed lines are the result of maximum mixing flavor
oscillation of the conventional atmospheric $\nu_{\mu}$ flux, for
$\Delta m^{2}$ around the Super-Kamiokande value (thick dot-dashed
line), as explained in the text. Solid, dotted, dashed and
long-dashed lines correspond to the upper bounds in Figure~1,
subjected to vacuum flavor oscillation, averaged in transit to
Earth. Fluxes are multiplied by $E^{2}$.}
\label{fig:2}
\end{figure}

The allowed band for atmospheric charm-induced tau-neutrinos is  
presented in Figure~\ref{fig:2}. In the absence of neutrino flavor  
oscillations, atmospheric $\nu_{\tau}$'s from charm are the only source  
of tau-neutrinos and the prompt window is completely open. We remind  
the reader that cosmic beam dumps produce negligible fluxes of  
tau-neutrinos. Recent measurements of the Super-Kamiokande  
experiment\cite{superk} do however favor non-vanishing neutrino masses  
and produce compelling evidence for $\nu_{\mu} \rightarrow \nu_{\tau}$  
flavor transitions, with maximum mixing angle ($\sin^ {2} 2\theta = 1$)  
and a square mass difference of $\Delta m^{2} =3.2 \times 10^{-3}$  
eV$^{2}$. We therefore consider a scenario where the conventional  
atmospheric $\nu_{\mu}$ flux in Figure~\ref{fig:1} is modified by this  
oscillation mechanism. In the calculation we consider the range  
$\log_{10} (\Delta m^{2}/$eV$^{2}) = 2.5 \pm 0.5$. Similarly, the  
bounds bracketing the extragalactic diffuse muon-neutrino flux can be  
translated into estimates of bounds on a cosmic tau-neutrino flux by  
adopting a typical value for the ratio of $\nu_{\tau}$ to $\nu_{\mu}$  
vacuum flavor conversion, over astronomical baselines, of  
$F_{\nu_{\tau}/\nu_{\mu}} =0.5$\cite{halzen:prd62}. The resulting  
fluxes are also shown in Figure~\ref{fig:2}. The appearance of a  
tau-neutrino below 1~TeV is a clear indication of atmospheric neutrino  
oscillations from a conventional $\nu_{\mu}$ flux. Above $\approx  
2$~TeV, we face again the interplay between extragalactic bounds and  
the atmospheric prompt component. For MPR scenarios, the prompt window  
may be closed; for WB-bounds the prompt tau-neutrino appearance may  
dominate extragalactic oscillation fluxes up to 50~TeV (for evolving  
sources), or even 500~TeV (for non-evolving sources).

In conclusion, we observe that the calculations of prompt neutrino  
fluxes are subjected to large uncertainties, arising from possible  
extreme cascade parameters combined with different charm production  
models. We identified the possibility that prompt fluxes are observable  
in a kilometer-scale neutrino telescope, particularly in the region  
from 2~TeV to 2~PeV. At the upper range of the predictions, a small  
component of prompt neutrinos may already be present in the observed  
sample of high-energy atmospheric neutrinos of the AMANDA  
experiment\cite{nature}.

Detection of the atmospheric prompt-$\nu_{\mu}$ will provide unique  
information on heavy quark interactions at energies not accessible to  
particle accelerators. The better characterization of the charm induced  
neutrino component will increase the discrimination power of neutrino  
detectors at higher energies. Knowledge of atmospheric charm production  
is essential in identifying the origin of a detected  
$\nu_{\tau}$ above a few TeV, whose appearance may also result from  
neutrino flavor oscillations.

%
\acknowledgements
We would like to thank Jean-Marie Fr\`ere for valuable discussions
and suggestions. This work was partially supported by the
I.I.S.N. (Belgium), by The Communaut\'e Fran\c{c}aise de
Belgique - Direction de la  Recherche Scientifique, programme
ARC, by the U.S.~Department of Energy under Grant No.~DE-FG02-95ER40896 and by the University of Wisconsin Research Committee with funds granted by the Wisconsin Alumni Research Foundation.


\mediumtext
%
\begin{table}  
\caption[]{\label{table:1} Upward-going muon and tau
event rates per year arising from charged current interactions of
atmospheric neutrinos in ice or water, for different neutrino
telescopes' effective areas and thresholds. 
Prompt flux calculation based on maximum (MAX) and minimum (MIN)
range limits described in the text.}
\smallskip
\centering
\begin{tabular}{lcc|lll|ll}
\hline
          &&& \multicolumn{3}{c}{$\mu^{+}+\mu^{-}$ }
           & \multicolumn{2}{c}{$\tau^{+}+\tau^{-}$}\\
Experiment & Threshold &$A_{eff}$ &Conventional    
& Prompt & Prompt & Prompt & Prompt \\
           &&&& MAX & MIN & MAX & MIN \\ \hline
BAIKAL NT-200 &1  TeV  &$2\times10^{3}$~m$^{2}$ 
&22    &3    &0   &0   &0 \\
AMANDA-II     &1  TeV  &$3\times10^{4}$~m$^{2}$ 
&330   &44   &2   &7   &0 \\
ICECUBE       &1  TeV  &1 km$^{2}$
&11000 &1470 &53  &216 &11 \\
ICECUBE       &10 TeV  &1 km$^{2}$
&170   &157  &3   &22  &1  \\
\hline
\end{tabular}
\end{table}

%
\newpage
\newpage
\end{document}